\documentclass[conference]{IEEEtran}
\IEEEoverridecommandlockouts
\usepackage{cite}
\usepackage{amsmath,amssymb,amsfonts}
\usepackage{algorithmic}
\usepackage{graphicx}
\usepackage{textcomp}
\usepackage{xcolor}
\usepackage{adjustbox}
\usepackage{multirow}
\usepackage{subcaption}
\usepackage{soul}
\usepackage[ruled,vlined]{algorithm2e}
\def\BibTeX{{\rm B\kern-.05em{\sc i\kern-.025em b}\kern-.08em
    T\kern-.1667em\lower.7ex\hbox{E}\kern-.125emX}}
\begin{document}

\author{\IEEEauthorblockN{Quincy Card, Kshitiz Aryal, Maanak Gupta}
\IEEEauthorblockA{\textit{Department of Computer Science} \\
\textit{Tennessee Tech University,}
Cookeville, TN USA\\
qacard42@tntech.edu, karyal42@tntech.edu, mgupta@tntech.edu}}

\title{Explainability-Informed Targeted \\ Malware Misclassification}

\maketitle

\begin{abstract}
In recent years, there has been a surge in malware attacks across critical infrastructures, requiring further research and development of appropriate response and remediation strategies in malware detection and classification. Several works have used machine learning models for malware classification into categories, and deep neural networks have shown promising results. However, these models have shown its vulnerabilities against intentionally crafted adversarial attacks, which yields misclassification of a malicious file. Our paper explores such adversarial vulnerabilities of neural network based malware classification system in the dynamic and online analysis environments. To evaluate our approach, we trained Feed Forward Neural Networks (FFNN) to classify malware categories based on features obtained from dynamic and online analysis environments. We use the state-of-the-art method, SHapley Additive exPlanations (SHAP), for the feature attribution for malware classification, to inform the adversarial attackers about the features with significant importance on classification decision. Using the explainability-informed features, we perform targeted misclassification adversarial white-box evasion attacks using the Fast Gradient Sign Method (FGSM) and Projected Gradient Descent (PGD) attacks against the trained classifier. Our results demonstrated high evasion rate for some instances of attacks, showing a clear vulnerability of a malware classifier for such attacks. We offer recommendations for a balanced approach and a benchmark for much-needed future research into evasion attacks against malware classifiers, and develop more robust and trustworthy solutions.
\end{abstract}


\section{Introduction}
Malware poses a significant cybersecurity threat, demanding an effective classification system for swift remediation. The process of addressing malware that has invaded a computer system broadly involves detecting it, classifying it, and then addressing the malware based on this classification. Detection refers to identifying the presence or absence of malware. Detection problems are sometimes referred to as binary classifications. This study distinguishes detection from classification, which aims to discern between various types of malware samples. Classification categorizes malware into families or categories, with ``family" referring to variants that share common characteristics, and ``category" grouping malware based on their objectives (e.g., ransomware encrypting systems or files).

Different types of malware necessitate tailored response plans, with adware requiring different treatment from trojans or ransomware. Machine learning-based classification methods have emerged to address this need \cite{tobiyama2016}. These methods are broadly categorized into static \cite{rahalilashkari2020}, dynamic \cite{keyes2021}, and online analysis \cite{kimmell2021analyzing}. Static analysis inspects dormant malicious files but is susceptible to obfuscation. Dynamic analysis executes malware in a controlled, simulated environment, while online analysis monitors systems in real time, preventing malware from detecting it is in a sandbox and remaining dormant. However, conducting online analysis can be resource-intensive, especially in preventing malware from accessing the internet in online environments.
Despite the importance of accurate malware classification, researchers often overlook the vulnerability of models to adversarial examples crafted to deceive trained models \cite{demetrio2019static, demetrio2020static, khormali2019static, YUSTE20221stativ,aryal2023exploiting, aryal2024intra, aryal2021survey}. Addressing this vulnerability is crucial for ensuring the reliability and robustness of classification systems, particularly in dynamic and online malware analysis where research is limited.

Explainable AI methods focus on feature attribution, elucidating model decisions and enhancing user trust. These explanations play a vital role in identifying crucial features for malware classification, empowering security analysts in countering threats. However, they can also inform the generation of adversarial evasion attacks as well \cite{demetrio2019explaining}. By indicating which features are important to model decision-making, an adversary is able to create adversarial samples that are more effective at fooling the classifier into misclassification, thereby intentionally evading proper classification.

White-box and black-box evasion attacks represent two distinct approaches to crafting adversarial examples to deceive machine learning models. In a white-box attack, the attacker has full access to the target model's architecture, parameters, and training data, enabling them to directly manipulate the model's input features to generate adversarial examples. Conversely, black-box attacks occur when the attacker has limited or no access to the target model's internal structure or training data. In such cases, the attacker interacts with the model by querying it with inputs and observing the corresponding outputs. Despite their differences, both types of attacks aim to exploit vulnerabilities in machine learning models to undermine their performance and reliability.

\begin{table*}[!t]
\caption{Related works compared. A $\surd$ indicates that a specific paper has this attribute, and a blank cell shows that this attribute or model does not exist.}
\label{tab:related-works}
\begin{adjustbox}{width=\textwidth}
\begin{tabular}{|l|c|ccc|ll|ll|}
\hline
\multicolumn{1}{|c|}{\multirow{2}{*}{Paper}} & \multirow{2}{*}{Target Model} & \multicolumn{3}{c|}{Analysis} & \multicolumn{2}{c|}{Domain} & \multicolumn{2}{c|}{Platform} \\ \cline{3-9} 
\multicolumn{1}{|c|}{} &  & \multicolumn{1}{c|}{Detection} & \multicolumn{1}{c|}{Category Classification} & \multicolumn{1}{c|}{Family Classification} & \multicolumn{1}{c|}{Dynamic} & \multicolumn{1}{c|}{Online} & \multicolumn{1}{c|}{Android} & \multicolumn{1}{c|}{Windows} \\ \hline
Stokes et al. (2018) \cite{stokes2018} & FFNN & \multicolumn{1}{c|}{$\surd$} & \multicolumn{1}{l|}{} &  & \multicolumn{1}{c|}{$\surd$} & & \multicolumn{1}{l|}{} & \multicolumn{1}{c|}{$\surd$}  \\ \hline
Kucuk et al. (2020) \cite{kucuk2020} & Random Forest & \multicolumn{1}{l|}{} & \multicolumn{1}{l|}{} & \multicolumn{1}{c|}{$\surd$} & \multicolumn{1}{c|}{$\surd$} & & \multicolumn{1}{l|}{} & \multicolumn{1}{c|}{$\surd$} \\ \hline
Ahmed et al. (2022) \cite{AHMED2022107903} & Ensemble & \multicolumn{1}{c|}{$\surd$} & \multicolumn{1}{l|}{} & & \multicolumn{1}{c|}{$\surd$}  & & \multicolumn{1}{c|}{$\surd$} & \\ \hline
Rafiq et al. (2023) \cite{rafiq2023} & AutoML & \multicolumn{1}{c|}{$\surd$} & \multicolumn{1}{l|}{} & & \multicolumn{1}{c|}{$\surd$} &  & \multicolumn{1}{c|}{$\surd$}  &  \\ \hline
\textbf{Our Approach} & FFNN & \multicolumn{1}{l|}{} & \multicolumn{1}{c|}{$\surd$} & & \multicolumn{1}{c|}{$\surd$} & \multicolumn{1}{c|}{$\surd$} & \multicolumn{1}{c|}{$\surd$}  & \multicolumn{1}{c|}{$\surd$} \\ \hline
\end{tabular}
\end{adjustbox}
\end{table*}

This paper utilizes SHapley Additive exPlanations (SHAP) to interpret black-box model decisions, informing targeted malware misclassification attacks. A Feed Forward Neural Network (FFNN) based malware classifier is trained for both dynamic and online malware datasets, with \textit{DeepSHAP} \cite{lundberg2017} providing global interpretations. These explanations informed the white-box Fast Gradient Sign Method (FGSM) and Projected Gradient Descent (PGD) attacks where we selected the Ransomware and Adware classes as our targets for the dynamic analysis and the Ransomware and PUA classes for the online analysis. In some instances of our attacks, we observed almost perfect evasion from the malware classifier, highlighting a clear vulnerability in this deep-learning model.

The main contributions of this work include:
\begin{itemize}
\item We evaluate the effectiveness of deep learning models for classifying malware categories in both dynamic and an online malware analysis data set.
\item We extend this analysis by explaining model predictions on a global level using  SHapley Additive exPlanations (SHAP).
\item We use these SHAP explanations to inform white-box evasion attacks on the deep learning models for targeted misclassification.
\end{itemize}
The paper is organized as follows. Section \ref{sec:background} reviews related works in evasion attacks conducted in the sphere of dynamic analysis. Section \ref{sec:methodology} outlines the methodology and introduces the dynamic and online datasets. Results, model explanations, and evasion attacks for each dataset are presented in Section \ref{sec:results}. The paper concludes with a summary and discussion of future work in Section \ref{sec:conclusion}.

\section{Related Work}\label{sec:background}
The field of adversarial attacks in dynamic malware analysis is relatively new in academia, with limited prior research available. Existing studies primarily focus on targeted misclassification, which involves deliberately choosing incorrect labels and crafting adversarial examples to deceive classifiers into classifying them as such. This is not the same as untargeted misclassifications, where adversarial examples are crafted to cause models to misclassify input without aiming at specific classes with the aim of decreasing accuracy and trust in a model's ability to classify correctly. Table \ref{tab:related-works} summarizes relevant work, outlining features such as the domain, type of analysis, targeted models, and malware platform.

For instance, \cite{stokes2018} utilized the Jacobian method to inform targeted misclassification attacks on an FFNN model detecting Windows malware. Similarly, \cite{AHMED2022107903} conducted evasion attacks on an ensemble model detecting Android ransomware, employing information gain to guide the attacks. \cite{rafiq2023} and \cite{kucuk2020} also explored evasion attacks on machine learning models detecting Android malware and Windows malware families, respectively, with the latter being one of the few to delve into targeted misclassification using a Random Forest model.

Notably, none of the mentioned works specifically address targeted misclassification on deep-learning classifier models for dynamic or online malware categories, highlighting a gap in the literature that our work aims to fill. We propose using black-box models to classify malware categories in dynamic and online datasets, followed by using the SHAP explainability method to guide adversarial attacks and deceive the model into misclassifying samples as targeted categories.

\section{Methodology}\label{sec:methodology}
Our methodology can be divided into three major parts: 1) Data collection for online and dynamic malware analysis; 2) Training the malware classifier for each dataset; and 3) Targeted adversarial attacks on the trained malware classifiers. This section discusses our methodology for training our dynamic analysis, online analysis classifier model, and approach to adversarial attacks for targeted misclassification.

\subsection{Dynamic Analysis}
For the dynamic malware analysis, we utilized the AndMal2020 dataset from the Canadian Center for Cybersecurity \cite{cccsc2020}, containing 12 different Android malware classes. Each sample comprises 141 features across six categories: \textit{memory},\textit{ API calls}, \textit{network}, \textit{battery}, \textit{log writing}, and \textit{total processes}. The class distribution in this dynamic malware dataset is highly imbalanced, as seen in Table \ref{tab:dynamic-distribution}. We attempted to address this imbalance by excluding minority classes and adjusting class weights; however, the SMOTE (Synthetic Minority Oversampling Technique) proved effective as it balanced the dataset by generating synthetic samples between real examples.
This resulted in 7261 samples per class, totalling 87132 samples. To train the classifier on this dynamic analysis dataset, we employed an FFNN, aiming for simplicity and establishing a baseline for future evasion attacks. The FFNN architecture included six hidden layers, two fully connected layers, and one dropout layer. The training was conducted on 80\% of the dataset for 135 epochs with a batch size of 10 while using the remaining 20\% for testing and validation. 

\begin{table}
    \caption{All Classes of Dynamic Data Set}
    \label{tab:dynamic-distribution}
    \centering
    \begin{tabular}{|c|c|}
    \hline
        Category & Number of Samples \\
        \hline
        Riskware & 7261 \\ \hline
        Adware & 5838 \\ \hline
        Trojan & 4412 \\ \hline
        Ransomware & 1861  \\ \hline
        Trojan\_Spy & 1801  \\ \hline
        Trojan\_SMS & 1028  \\ \hline
        Trojan\_Dropper & 837 \\ \hline
        PUA & 665  \\ \hline
        Backdoor & 591  \\ \hline
        Scareware & 462 \\ \hline
        FileInfector & 129  \\ \hline
        Trojan\_Banker & 118 \\
        \hline
    \end{tabular}
\end{table}

\subsection{Online Analysis}
For online analysis, we utilized the RaDaR dataset from the Indian Institute Technology Madras \cite{karapoola2022}, capturing Windows malware behavior on a real-time physical testbed. This dataset enables analysis of modern malware that is capable of detecting sandbox environments and remaining dormant. However, this level of analysis requires significant resources compared to dynamic analysis, with increased computation times due to larger data volumes. This dataset comprises five malware classes and 55 hardware-level features. To address the class imbalance evident in Table \ref{tab:online-distribution}, SMOTE was employed, resulting in 158158 samples per class, totalling 790790 samples. While we initially considered Long Short-Term Memory (LSTM) models for their time-series data advantages, we opted for FFNNs to maintain consistency across analyses. The FFNN architecture included five hidden layers with \textit{ReLU} activation, two fully connected layers, one dropout layer, and \textit{Softmax} activation for the output layer. Training utilized 80\% of the data for 100 epochs with a batch size of 50, with testing on the remaining 20\%.


\begin{table}
    \caption{All Classes of Online Data Set}
    \label{tab:online-distribution}
    \centering
    \begin{tabular}{|c|c|}
    \hline
        Category & Number of Snapshots \\
        \hline
        Cryptominer & 158158 \\ \hline
        Deceptor & 99099 \\ \hline
        Ransomware & 13013 \\ \hline
        PUA & 3003  \\ \hline
        Backdoor & 1001 \\
        \hline
    \end{tabular}
\end{table}

\subsection{Adversarial Approach}

For targeted adversarial attacks on the trained classifiers, we chose the Fast Gradient Sign Method (FGSM) \cite{goodfellow2015explaining} and Projected Gradient Descent (PGD) \cite{madry2019deep} attacks, both white-box attacks requiring knowledge of the target model architecture. These attacks were conducted as benchmarks for future research in dynamic and online malware classification, addressing the lack of existing work in targeted misclassification of black-box classified dynamic and online malware samples. Our attacks aim to generate adversarial examples that are misclassified as specific malware categories to assess the robustness of the models for classifying a particular malware class.

To inform the attacks, we first needed to have SHAP identify which features were important to model decision-making. As opposed to the related works that focused on targeting a transparent model, our work targeted a black-box model that needs post-hoc explanations to identify which features would be most effective in perturbing. We used SHAP's \textit{DeepExplainer} \cite{lundberg2017} to compute the SHAP values we used to inform the attacks. We used a subsample of 1000 samples from the dynamic test data set and 10000 samples from the online test data set. This under-sampling is necessary due to the resource-intensive nature of computing SHAP values balanced with the need to create effective adversarial examples. Our work thus properly balances the level of analysis achieved from the online analysis with the more resource-efficient dynamic analysis.

We performed a grid search to find the optimal set of hyperparameters for our misclassification attack. For adversarial evasion attacks in dynamic analysis, optimal hyperparameters for FGSM included an epsilon ($\epsilon$) of \textit{1.0} and a step size of \textit{0.8} with the \textit{L2 norm} bound on the perturbation. At the same time, optimal hyperparameters for PGD were an epsilon ($\epsilon$) of \textit{1.0}, step size of\textit{ 0.8}, and maximum iterations of \textit{50} with the \textit{L-Infinity norm} bound for perturbation. In online analysis, optimal FGSM parameters were epsilon \textit{1.0} and step size \textit{0.5} with the\textit{ L2 norm} bound, while optimal PGD parameters were epsilon \textit{1.0}, step size \textit{0.5}, and maximum iterations \textit{50} with the \textit{L-infinity norm} bound. The evaluation focused on the success rate of attacks, indicating the ratio of successfully misclassified examples to total generated examples. Our methodology for generating adversarial examples and assessing their success is outlined in Algorithm \ref{alg:attack}.

\begin{algorithm}[!t]
  \caption{Algorithm for Adversarial Attack}
  \label{alg:attack}
  \SetAlgoLined
  \SetKwInOut{Input}{Input}
  \SetKwInOut{Output}{Output}
  
  \Input{Classifier model $C$, test malware data $X$, number of features to perturb $num\_features$, SHAP values $shap\_values$ }
  \Output{Attack success rate for FGSM and PGD}
  
  \textbf{1.} Wrap the Classifier model $C$ with an ART estimator\;
  \textbf{2.} Configure attack parameters for targeted misclassification\;
  \textbf{3.} Specify the target class for each adversarial example\;
  \textbf{4.} Identify the most important features from $shap\_values$\;
  \textbf{5.} Generate adversarial examples using FGSM and PGD\;
  \textbf{6.} Evaluate the classifier on adversarial examples\;
  \textbf{7.} Calculate success rates of evasion for FGSM and PGD\;
\end{algorithm}

\begin{figure*}
  \centering

  \subfloat[Confusion Matrix for Dynamic Data set]{\includegraphics[width=0.5\textwidth]{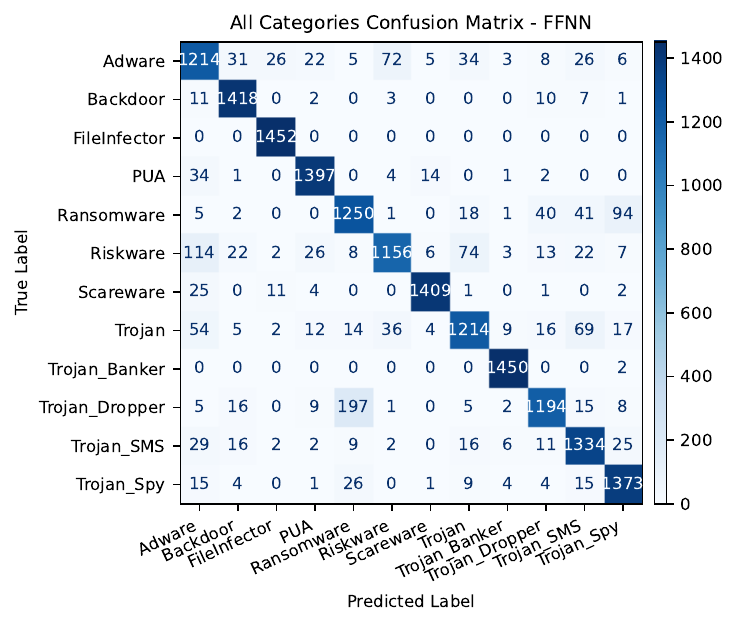}}\hfill
  \subfloat[Confusion Matrix for Online Data set]{\includegraphics[width=0.4\textwidth]{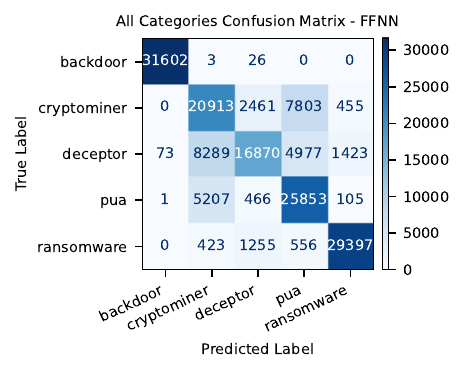}}

  \caption{Performance of models in Dynamic and Online Analysis}
  \label{fig:all_categories_cm}
\end{figure*}

\begin{table}[]
\caption{Performance Metrics for Dynamic Analysis and Class-Specific Metrics}
\label{tab:dynamic-performance}
\begin{adjustbox}{width=0.47\textwidth}
\begin{tabular}{|c|c|c|c|c|}
\hline
                            & Accuracy (\%) & Precision (\%) & Recall (\%) & F1 (\%) \\ \hline
FFNN without SMOTE          & 81.52         & 81.63          & 81.52       & 81.58   \\ \hline
FFNN with SMOTE             & 91.01         & 91.06          & 91.01       & 91.03   \\ \hline
Ransomware with Smote       & 86.09         & 82.84          & 86.09       & 84.43   \\ \hline
Adware with Smote           & 83.61         & 80.61          & 83.61       & 82.08   \\ \hline
\end{tabular}
\end{adjustbox}
\end{table}

\begin{table}[]
\caption{Performance Metrics for Online Analysis and Class-Specific Metrics}
\label{tab:online-performance}
\begin{adjustbox}{width=0.46\textwidth}
\begin{tabular}{|c|c|c|c|c|}
\hline
                          & Accuracy (\%) & Precision (\%) & Recall (\%) & F1 (\%) \\ \hline
FFNN without SMOTE        & 77.59         & 78.37          & 77.59       & 77.98   \\ \hline
FFNN with SMOTE           & 78.55         & 79.94          & 78.55       & 79.24   \\ \hline
Ransomware with Smote     & 92.88         & 100.00         & 92.88       & 96.31   \\ \hline
PUA with Smote            & 84.07         & 100.00         & 84.07       & 91.35   \\ \hline
\end{tabular}
\end{adjustbox}
\end{table}

\section{Results and Discussion}\label{sec:results}

\subsection{Evaluation of Performance Metrics for classifier models}
The first part of our work is to train the classifier models on dynamic and online datasets. Table \ref{tab:dynamic-performance} provides performance metrics for overall model performance in dynamic analysis and for our target classes for attack: Ransomware and Adware. The comparison with and without SMOTE intervention indicates a notable performance increase with synthetic samples. The F1-score close to 91\% for overall performance in Table \ref{tab:dynamic-performance} proves FFNN works well for all our scenarios. Despite Ransomware being a top majority class in the unaltered dataset, its relatively poor performance suggests potential overlap between SMOTE's synthetic samples for minority classes and the decision boundary of majority classes, but we have concluded that using SMOTE is still viable due to the increased overall classifier performance.

Performance metrics for overall model performance for the online analysis before and after SMOTE intervention can be found in Table \ref{tab:online-performance}. Ransomware and PUA are our target classes for attacks, and the classifier's performance in classifying these two classes is shown in Table \ref{tab:online-performance}. The FFNN model had an F1 score of 79.24\%. This was likely because of not maintaining the time-series nature of the data and because of SMOTE's synthetic samples, as evidenced by the majority classes being outperformed by the minority classes in Figure \ref{fig:all_categories_cm}. A marginal increase in performance is observed when using SMOTE. This decision to use SMOTE was motivated by the desire to maintain consistency with the methodology employed for dynamic analysis. Consequently, both the model and the dataset influenced by SMOTE were utilized for the generation of explanations and adversarial examples. This decision also aligns with how real-world professionals may choose to manage the imbalanced class distribution. After training the model, we explained the model's predictions by applying SHAP.

\begin{figure*}
  \centering

  \subfloat[Dynamic]{\includegraphics[width=0.5\textwidth]{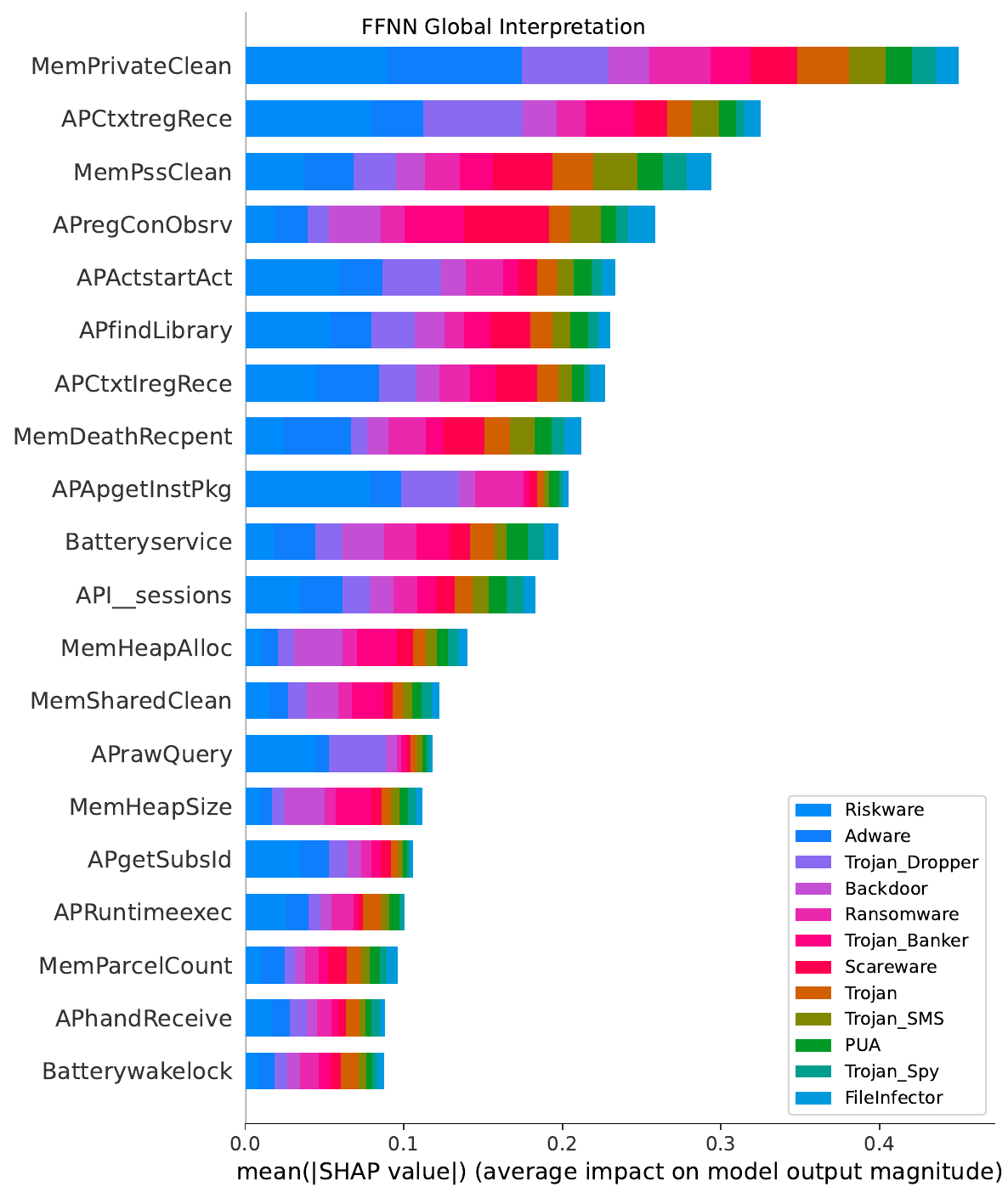}}\hfill
  \subfloat[Online]{\includegraphics[width=0.5\textwidth]{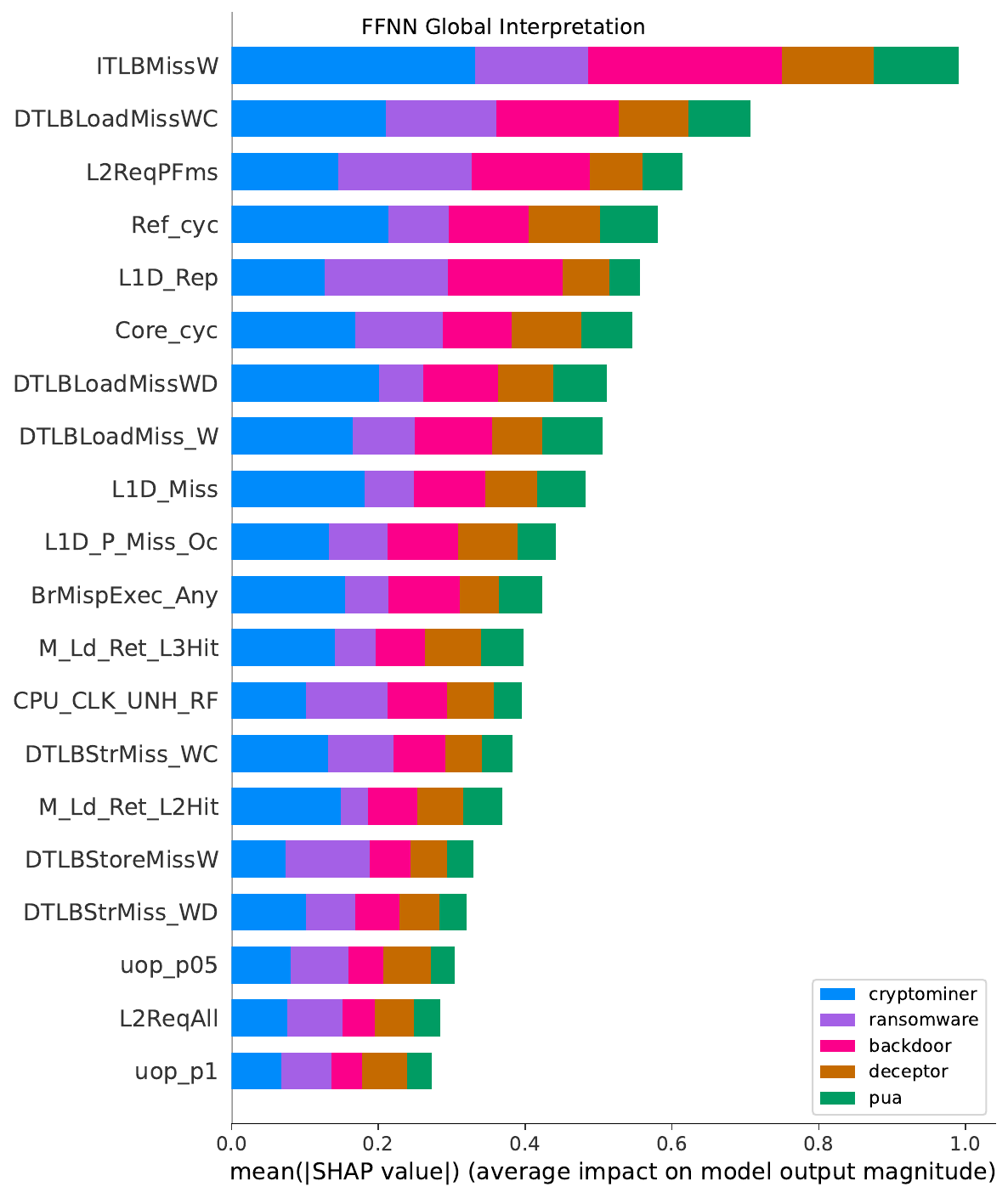}}

  \caption{A stacked bar graph depicting the top 20 online features identified by SHAP in model decision making}
  \label{fig:global_shap}
\end{figure*}

\begin{figure*}
  \centering

  \subfloat{\includegraphics[width=0.5\textwidth]{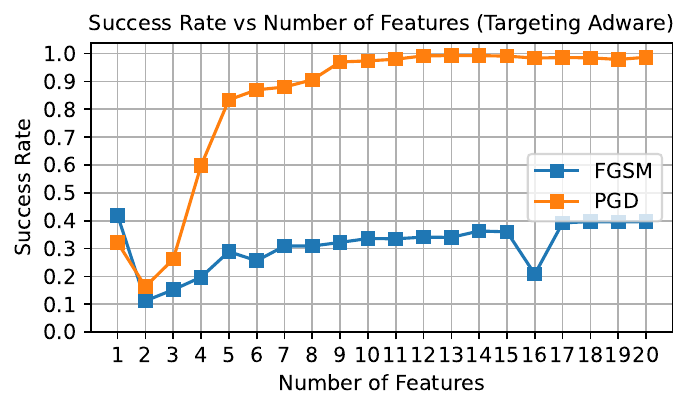}}\hfill
  \subfloat{\includegraphics[width=0.5\textwidth]{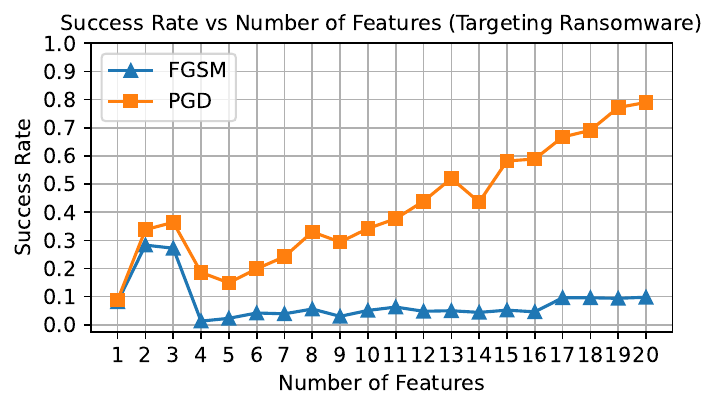}}

  \caption{Success rates for FGSM and PGD attacks in targeting dynamic Adware and Ransomware categories}
  \label{fig:dynamic_success_rates}
\end{figure*}

\begin{figure*}
  \centering

  \subfloat{\includegraphics[width=0.5\textwidth]{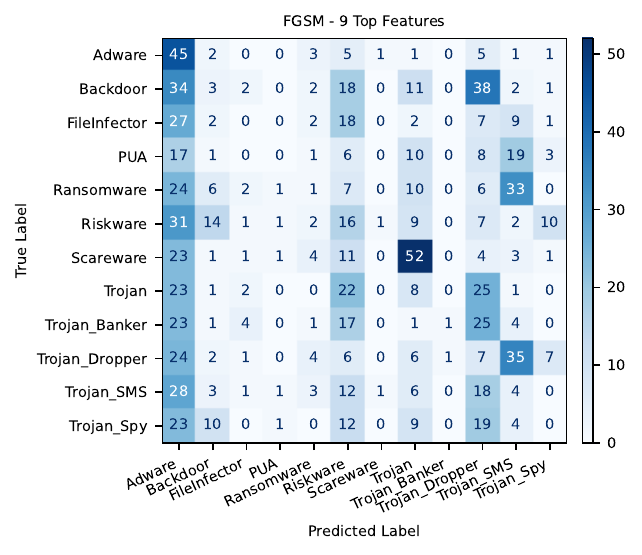}}\hfill
  \subfloat{\includegraphics[width=0.5\textwidth]{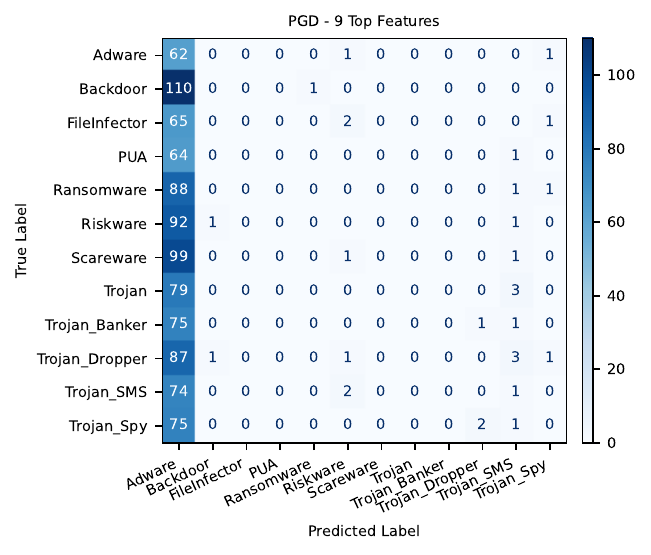}}

  \caption{Confusion matrices for FGSM and PGD attacks in targeting dynamic Adware}
  \label{fig:dynamic_adware_conf}
\end{figure*}

\begin{figure*}
  \centering

  \subfloat{\includegraphics[width=0.5\textwidth]{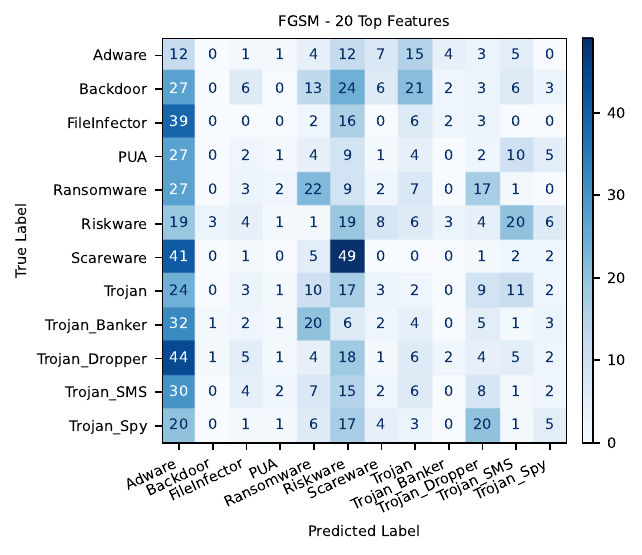}}\hfill
  \subfloat{\includegraphics[width=0.5\textwidth]{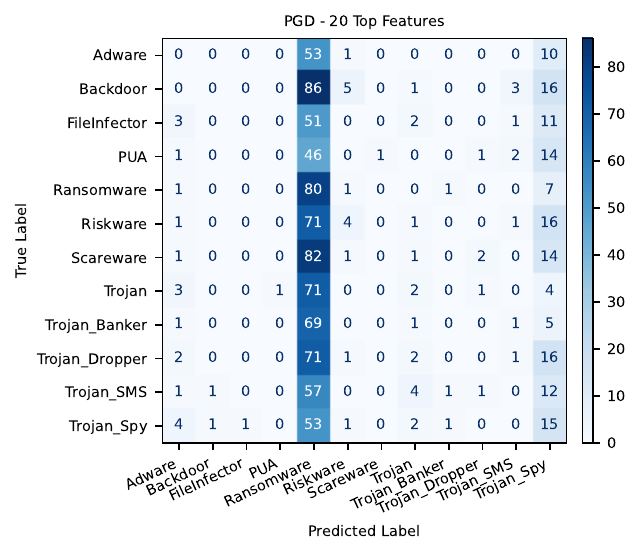}}

  \caption{Confusion matrices for FGSM and PGD attacks in targeting dynamic Ransomware}
  \label{fig:dynamic_ransomware_conf}
\end{figure*}

\begin{figure*}
  \centering

  \subfloat{\includegraphics[width=0.5\textwidth]{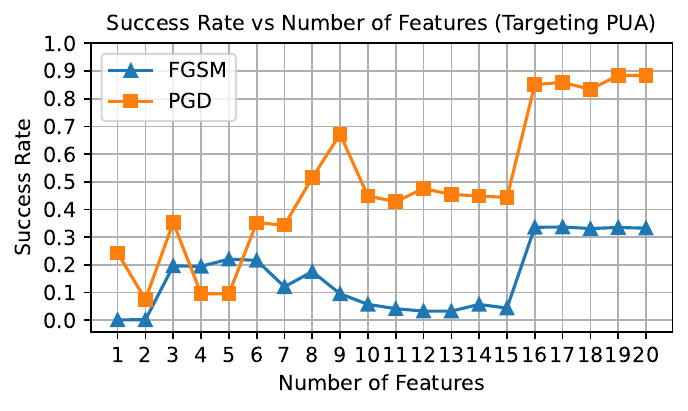}}\hfill
  \subfloat{\includegraphics[width=0.5\textwidth]{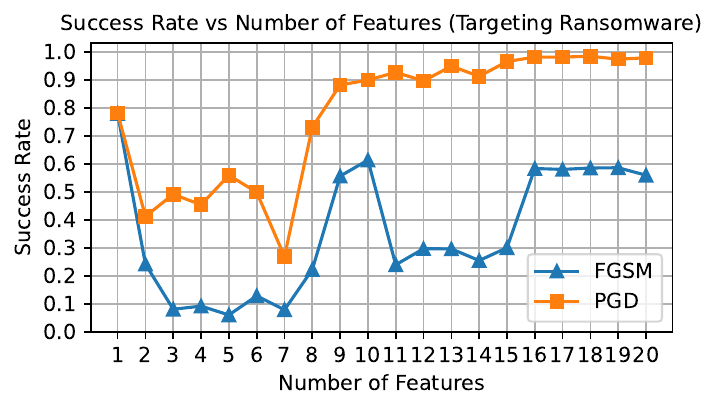}}

  \caption{Success rates for FGSM and PGD attacks in targeting online PUA and Ransomware categories}
  \label{fig:online_success_rates}
\end{figure*}

\begin{figure*}
  \centering

  \subfloat{\includegraphics[width=0.4\textwidth]{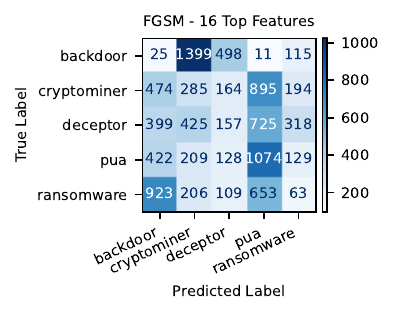}}\hfill
  \subfloat{\includegraphics[width=0.4\textwidth]{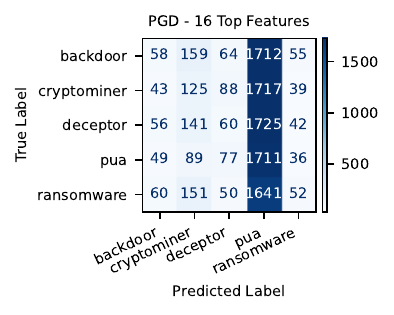}}

  \caption{Confusion matrices for FGSM and PGD attacks in targeting online PUA}
  \label{fig:online_pua_conf}
\end{figure*}

\begin{figure*}
  \centering

  \subfloat{\includegraphics[width=0.4\textwidth]{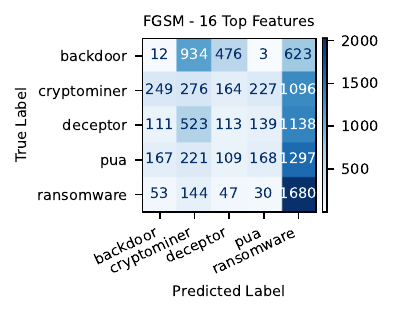}}\hfill
  \subfloat{\includegraphics[width=0.4\textwidth]{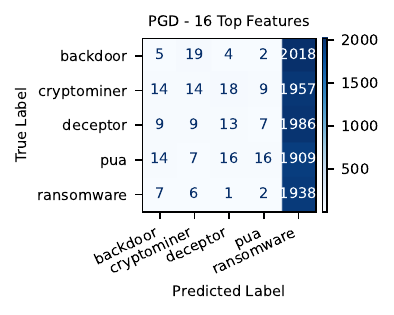}}

  \caption{Confusion matrices for FGSM and PGD attacks in targeting online Ransomware}
  \label{fig:online_ransomware_conf}
\end{figure*}

\subsection{Global Explanation}
With a reduced sample size, SHAP's {\itshape DeepExplainer} computed SHAP values in 70 seconds the dynamic data set and about 15 minutes for the online data set. The summary plots in Figure \ref{fig:global_shap} illustrate feature importance across categories for both data sets. The x-axis shows the average magnitude impact on model output, with features arranged by their effects' magnitudes. Notably, the top features for the model trained on the dynamic data set were mostly API calls or Memory features, emphasizing their significance. We used the top 20 identified features for each respective data set to inform the targeted evasion attacks. We selected the Adware and Ransomware categories for the dynamic analysis and the PUA and Ransomware categories for the online analysis because they pose the highest potential damage if an adversary successfully misclassifies their malware. Misclassification as a less or more dangerous category than a sample's ground truth could significantly undermine the effectiveness of the remediation plan.

\subsection{Targeted Misclassification}
For the dynamic analysis, Figure \ref{fig:dynamic_success_rates} shows the rate of adversarial examples that were successfully able to fool the classifier. On targeting Adware, the PGD attack reached an acceptable success rate at just 9 out of 141 features, while the success rate of the FGSM attack peaked at around 40\% at 17 features being perturbed. For targeting Ransomware, the PGD attack required all 20 features to be perturbed to reach a success rate of 80\%, while the FGSM attack peaked at just above 30\% and then degraded in performance drastically, making this category comparatively worse at being targeted than Adware since it took more features for PGD and the FGSM's peak success rate was less. When comparing the attacks just targeting the Ransomware class, the FGSM attack vastly under-performs compared to the PGD attack, and Figure \ref{fig:dynamic_ransomware_conf} shows that this may be because it's targeting the incorrect class. For an untargeted misclassification problem, these results would still be promising as the model's ability to correctly classify any samples has greatly decreased, as seen in how the expected darker colouring of the diagonal in Figure \ref{fig:all_categories_cm} is not present at all, indicating the classifier's decreased ability to correctly predict a sample's true label. This work, however, is concerned with successful targeting of a selected class. The more complex, iterative PGD attack is capable of targeted misclassification, as Figure \ref{fig:dynamic_ransomware_conf} shows a distinct, vertical line on the "predicted Ransomware" column. There's a less dark vertical line on the "predicted Trojan\_Spy" column, making this the reason why the the PGD attack targeting Ransomware had a poorer performance than the PGD attack targeting the Adware. 

Similar to the attacks targeting the Ransomware category, the PGD attack when targeting the Adware class is much more effective than the FGSM attack, with Figure \ref{fig:dynamic_adware_conf} revealing this is also due to an error in targeting the incorrect class. The "predicted Adware" column shows that this FGSM attack is somewhat working, but it is not overwhelmingly effective like the PGD attack. We hypothesize that if more features were selected to be perturbed the PGD attack targeting Ransomware could at some point reach a comparable success rate as the one targeting Adware, as seen in the general upward trend of the success rate. However, we believe perturbing any more than 20 features could affect the malware's ability to function. Future work should investigate this.

For the online analysis, Figure \ref{fig:online_success_rates} shows the rate of adversarial examples that were successfully able to fool the classifier. Similar to the dynamic analysis, for both targeted misclassifications, the more complex iterative PGD attack was more effective in creating examples that fooled the classifier than the simpler FGSM attack. Unlike the dynamic analysis, the success rate fluctuates in a more volatile fashion for the first few features. For targeting both selected categories, the PGD attack reached an acceptable success rate around 16 out of 55 features, with the FGSM attack doing better in successfully targeting Ransomware over PUA. Figures \ref{fig:online_pua_conf} and \ref{fig:online_ransomware_conf} show the confusion matrices generated by the attacks for the feature amount that resulted in the highest success rate, revealing that for both the PUA and the Ransomware targeted attacks using FGSM the adversarial examples seem to be properly targeting the respective categories, but it's not as effective as the PGD attacks. These FGSM attacks would be effective in untargeted misclassification as these Figures show the diagonal from Figure \ref{fig:all_categories_cm} has been disrupted. It should be noted that the PGD attacks show a lot of misclassification of samples as other classes. Perhaps this is due to inefficiencies within the classifier, as seen how there were so many samples in Figure \ref{fig:all_categories_cm} were already misclassified.

Comparing the two levels of analysis, we have concluded that it is easier to create successful and properly effective targeted misclassification attacks for the dynamic model that did well, than it is for the online model which had trouble properly categorizing the malware from the beginning. The success rates for the attacks on the dynamic model were consistently less volatile than the attacks on the online model. These effective attacks were also the result of using considerably less samples to train the attacks for the dynamic model than what was necessary for the online model. The attacks on the online model were more successful in producing untargeted misclassification. This is noteworthy and should be investigated further.

\section{Conclusion and Future Work}\label{sec:conclusion}
In this paper, we trained an FFNN malware classifier model for a dynamic and online analysis data set to classify malware categories. The imbalanced class distribution of dynamic dataset was partially overcome with the use of SMOTE, which slightly degraded model performance but not to an unacceptable degree. Future authors can also extend this work by investigating models that are adept at handling time-series data, explaining those models with methods more well-suited to time-series data, and investigating the performance of these models on different data sets. We used SHAP to explain these black-box models and then used those explanations to inform targeted misclassification white-box evasion attacks. We performed evasion attacks by targeting 3 different malware classes in particular: Ransomware, Adware and PUA, while using FGSM and PGD as algorithms for crafting attacks. We compared the performance of models on performing attacks different malware classes using the different number of features informed by SHAP. Our results showed the success rate of targeted misclassification attack is close to 100\% in some of the attack instances, demonstrating the serious vulnerability of the classifier model.

In future research, we will extend this work by launching black-box attacks to generate adversarial examples, and map the changes made to the features to an actual malware. Future work should also investigate the possibility and use cases of untargeted misclassification, since the FGSM attacks we devised for this work were so effective in degrading model performance. Our findings can also contribute to creating more robust models through adversarial training, which would enhance real-world malware remediation, aiding cyber-analysts and strategists in efficiently categorizing and coordinating responses to malware threats. 

\section*{Acknowledgment}
This work was supported by the NSF Scholarship for Service Program Award 2043324 and 2230609.

\bibliographystyle{IEEEtran}
\bibliography{paper}

\begin{thebibliography}{10}
\providecommand{\url}[1]{#1}
\csname url@samestyle\endcsname
\providecommand{\newblock}{\relax}
\providecommand{\bibinfo}[2]{#2}
\providecommand{\BIBentrySTDinterwordspacing}{\spaceskip=0pt\relax}
\providecommand{\BIBentryALTinterwordstretchfactor}{4}
\providecommand{\BIBentryALTinterwordspacing}{\spaceskip=\fontdimen2\font plus
\BIBentryALTinterwordstretchfactor\fontdimen3\font minus \fontdimen4\font\relax}
\providecommand{\BIBforeignlanguage}[2]{{%
\expandafter\ifx\csname l@#1\endcsname\relax
\typeout{** WARNING: IEEEtran.bst: No hyphenation pattern has been}%
\typeout{** loaded for the language `#1'. Using the pattern for}%
\typeout{** the default language instead.}%
\else
\language=\csname l@#1\endcsname
\fi
#2}}
\providecommand{\BIBdecl}{\relax}
\BIBdecl

\bibitem{tobiyama2016}
S.~Tobiyama, Y.~Yamaguchi, H.~Shimada, T.~Ikuse, and T.~Yagi, ``Malware detection with deep neural network using process behavior,'' in \emph{2016 IEEE 40th Annual Computer Software and Applications Conference (COMPSAC)}, vol.~2, 2016, pp. 577--582.

\bibitem{rahalilashkari2020}
A.~Rahali and et~al., ``Didroid: Android malware classification and characterization using deep image learning,'' in \emph{10th International Conference on Communication and Network Security (ICCNS2020)}, Tokyo, Japan, November 2020, pp. 70--82.

\bibitem{keyes2021}
D.~S. Keyes, B.~Li, G.~Kaur, A.~H. Lashkari \emph{et~al.}, ``Entroplyzer: Android malware classification and characterization using entropy analysis of dynamic characteristics,'' in \emph{2021 Reconciling Data Analytics, Automation, Privacy, and Security: A Big Data Challenge (RDAAPS)}, 2021.

\bibitem{kimmell2021analyzing}
J.~C. Kimmell, M.~Abdelsalam, and M.~Gupta, ``Analyzing machine learning approaches for online malware detection in cloud,'' in \emph{IEEE conference on smart computing (SMARTCOMP)}, 2021.

\bibitem{demetrio2019static}
L.~Demetrio and et~al., ``Explaining vulnerabilities of deep learning to adversarial malware binaries,'' \emph{arXiv preprint arXiv:1901.03583}, 2019.

\bibitem{demetrio2020static}
------, ``Efficient black-box optimization of adversarial windows malware with constrained manipulations,'' \emph{arXiv preprint arXiv:2003.13526}, 2020.

\bibitem{khormali2019static}
A.~Khormali and et~al., ``Copycat: practical adversarial attacks on visualization-based malware detection,'' \emph{arXiv preprint arXiv:1909.09735}, 2019.

\bibitem{YUSTE20221stativ}
J.~Yuste, E.~G. Pardo, and J.~Tapiador, ``Optimization of code caves in malware binaries to evade machine learning detectors,'' \emph{Computers \& Security}, vol. 116, p. 102643, 2022.

\bibitem{aryal2023exploiting}
K.~Aryal, M.~Gupta, and M.~Abdelsalam, ``Exploiting windows pe structure for adversarial malware evasion attacks,'' in \emph{Proceedings of the Thirteenth ACM Conference on Data and Application Security and Privacy}, 2023, pp. 279--281.

\bibitem{aryal2024intra}
K.~Aryal, M.~Gupta, M.~Abdelsalam, and M.~Saleh, ``Intra-section code cave injection for adversarial evasion attacks on windows pe malware file,'' \emph{arXiv preprint arXiv:2403.06428}, 2024.

\bibitem{aryal2021survey}
K.~Aryal, M.~Gupta, and M.~Abdelsalam, ``A survey on adversarial attacks for malware analysis,'' \emph{arXiv preprint arXiv:2111.08223}, 2021.

\bibitem{demetrio2019explaining}
L.~Demetrio, B.~Biggio, G.~Lagorio, F.~Roli, and A.~Armando, ``Explaining vulnerabilities of deep learning to adversarial malware binaries,'' \emph{arXiv preprint arXiv:1901.03583}, 2019.

\bibitem{stokes2018}
J.~W. Stokes and et~al., ``Attack and defense of dynamic analysis-based, adversarial neural malware detection models,'' in \emph{MILCOM 2018 - 2018 IEEE Military Communications Conference (MILCOM)}, 2018, pp. 1--8.

\bibitem{kucuk2020}
Y.~Kucuk and G.~Yan, ``Deceiving portable executable malware classifiers into targeted misclassification with practical adversarial examples,'' in \emph{Proceedings of the tenth ACM conference on data and application security and privacy}, 2020, pp. 341--352.

\bibitem{AHMED2022107903}
U.~Ahmed, J.~C.-W. Lin, and G.~Srivastava, ``Mitigating adversarial evasion attacks of ransomware using ensemble learning,'' \emph{Computers and Electrical Engineering}, vol. 100, p. 107903, 2022.

\bibitem{rafiq2023}
H.~Rafiq and et~al., ``Mitigating malicious adversaries evasion attacks in industrial internet of things,'' \emph{IEEE Transactions on Industrial Informatics}, vol.~19, no.~1, pp. 960--968, 2023.

\bibitem{lundberg2017}
S.~M. Lundberg and S.-I. Lee, ``A unified approach to interpreting model predictions,'' in \emph{Advances in Neural Information Processing Systems}, vol.~30, 2017.

\bibitem{cccsc2020}
\BIBentryALTinterwordspacing
C.~C. for Cyber~Security. (2020) {CCCS-CIC-AndMal2020}. [Online]. Available: \url{https://www.unb.ca/cic/datasets/andmal2020.html}
\BIBentrySTDinterwordspacing

\bibitem{karapoola2022}
S.~Karapoola, N.~Singh, C.~Rebeiro, and K.~V., ``Radar: A real-world dataset for ai powered run-time detection of cyber-attacks,'' in \emph{Proceedings of the 31st ACM International Conference on Information \& Knowledge Management}, 2022, pp. 3222--3232.

\bibitem{goodfellow2015explaining}
I.~J. Goodfellow, J.~Shlens, and C.~Szegedy, ``Explaining and harnessing adversarial examples,'' 2015.

\bibitem{madry2019deep}
A.~Madry, A.~Makelov, L.~Schmidt, D.~Tsipras, and A.~Vladu, ``Towards deep learning models resistant to adversarial attacks,'' 2019.

\end{thebibliography}

\end{document}